# Revision of Analytical Properties of Reaction Amplitude near Thresholds Using the Example of Muon-Induced Prompt Fission

F.F. Karpeshin

*D.I. Mendeleev All-Russian Research Institute for Metrology, 190005, Saint-Petersburg, Russia*

E-mail: fkarpeshin@gmail.com

Experimental data on the muon-induced prompt fission are analyzed and compared to theory. Good agreement is observed for the 2$p$–1$s$ and 3$d$–1$s$ transitions. Anomalously big nonradiative width in the case of 3$d$–1$s$ transition is surprised. Ways of resolving this puzzle are discussed.

## 1. Introduction

I start with an example which clearly demonstrates bounds of our knowledge. Consider fission of $^{238}$U on the final stage of saddle-to-scission descent. Two scenarios are possible. Neck may rupture at small distance between the fragments. Strong Coulomb repulsion accelerates the fragments to some asymptotic value of TKE. Otherwise, the neck may rupture later. The distance between the fragments is larger, Coulomb repulsion is weaker. But if the fragments already have an initial velocity, they can accelerate to the same TKE. So, in the first case, the distance is small, repulsion – great. In the second case, the distance is larger, repulsion is weaker, but the TKE are the same, however. The first scenario implies strong dissipation, the second – weak dissipation. Therefore, one can say nothing about dissipation basing on experiment.

However, there is a tool which can help. This is prompt fission in muonic atoms, induced by nonradiative muon transitions. And the muons play the role of Maxwell's demons, or spectators, who watch the fission spectacle directly from the inside, and transfer the information to us. The process is as follows.

Muons "slow down" in matter then they start to be captured into atoms, to the muonic orbits with n about 14. Then they cascade down by means of Auger- or radiative transitions. In the final transition, for example 2$p$ to 1$s$, they have a chance to transfer the energy to the nucleus. The nucleus gets excited, and then it can undergo fission, which is called prompt. Otherwise, the muon will be captured by the nucleus due to weak interactions, and the nucleus also can undergo fission. This is called delayed fission.

It was Wheeler who proposed such a process [1]. He supposed tat nuclear excitation can occur in the 2$s$–1$s$ transition. These transitions compete with the radiative 2$s$–2$p$ transitions. Other possible nonradiative transitions, like 2$p$–1$s$, 3$p$–1$s$ compete with the same radiative transitions. And 3$p$ to 1$s$ transition also competes with the radiative 3$p$–2$s$ transition. However, Zaretsky showed [2] that the probability of nuclear excitation in these transitions is also high, of the order of 1. Furthermore, Teller and Weiss showed [3] that in the $E$2 transition 3$d$–1$s$ the nonradiative probability is about the same. Those results were confirmed by Nesterenko and me [4]. Moreover, they showed that the nonradiative probability in the $E$3 transition, 3$d$–2$p$, is also high. It achieves 15% in spite of it competes with the $E$1 radiative transition. In this transition, excitation of the low-lying electric octupole resonance (LEOR) takes place. The energy 3.5 MeV is too small in order to induce fission, but LEOR can be studied in this way as excited by monochromatic photons.

Experimental discovery was induced due to Zaretsky's calculations. It took place in Dubna in 1960, with the paper by Pontekorvo et al. [5, 6]. For this purpose, muonic spectra of two atoms were compared to one another: $^{238}$U and $^{208}$Pb. They might be similar as uranium and lead have close atomic numbers. But the lead nucleus has a very low level density, which makes nonradiative excitation unlikely. Opposite, the level density is high in the U case, hence nonradiative excitation was expected.

When the spectra had been obtained, they both showed distinctive peaks corresponding to the 2 to 1 transition for both nuclei. In the case of U, the peak was by about 20% lower. Therefore, a missing intensity was established, which evidenced an appreciable probability of the nonradiative transition in uranium nuclei. This discovery was included in the register of scientific discoveries of the USSR.

Numeric data on the nonradiative probabilities were obtained in the group of Prof. David at the University of Bonn. They studied balance of incoming and outgoing radiative intensities for each atomic level. As a result, they obtained the nonradiative probabilities presented in Table 1. The data are compared with theoretical calculations. In particular, the data are shown for various components of the $2p$–$1s$ transition. The mechanism of radiationless transitions is considered to be as follows. Let muon make a transition e.g. from the $2p$ to $1s$ state. The energy of the transition is transferred to the nucleus by means of a virtual photon. This is actually a reverse internal conversion (IC) process. Theory of IC teaches that its probability factorizes into the nuclear radiation probability and ICC (Internal conversion coefficient), which is practically independent of the nuclear model.

In a number of theoretical papers calculations of the probabilities of the dipole $2p$, $3p$ → $1s$ as well as the quadrupole $3d$ → $1s$ radiationless transitions were carried out. Along with those transitions, herein we also consider the $3d$ → $2p$ radiationless transitions of $E$3 type accompanied by the excitation of the low-lying electric octupole giant resonance (LEOR). We shall show that they are expected to be of the same order of magnitude as the dipole and quadrupole transitions. For present purposes we have used the quasi-particle-phonon nuclear model (QPNM) [8], which is known to work well in the description of low-lying nuclear states as well as giant resonances both in spherical and deformed nuclei.

These results are listed in Table 1. They show a rather large variability of the radiative widths for various fine-structure components. Data for the $2p$–$1s$ transition are in agreement with theoretical calculations. The authors did not search for the $3d$–$2p$ transitions, as they did not know about Ref. [4] at the time of experiment. Surprisingly, a radical disagreement with theory was found for the $3p$–$1s$ transitions. It is not merely 20%. This difference shows that the ratio of nonradiative-to-radiative probabilities achieves an order of magnitude.

In order to better understand the nature of this divergence, I undertook model-independent calculation using the experimental cross-sections of photoexcitation and photofission of the uranium nuclei. This paper provides a detailed information concerning the cross-sections. Before turning to these results let us remind the physical principles which underlie the microscopic calculation in Ref. [4].

**Outline of the model**

Description of the radiationless transitions first of all requires knowledge of the nuclear electromagnetic strength functions which contain information about the nuclear structure. The strength functions from QPNM are quite appropriate for this purpose. Next important moment is the muonic conversion coefficients (MCC). Involving MCC makes the calculations considerably easier, reducing the problem to independent calculations of the nuclear strength functions and MCC due to factorization of the amplitude. Moreover, factorization justifies use of experimental cross-sections for calculation of the radiationless transition probabilities.

The expression for the radiationless transition width is obtained by using the principle of detailed balance. In accordance with the definition of the MCC the width of the inverse process is the product of the radiative nuclear width and the MCC only. Turning to the radiationless nuclear excitation one immediately obtains an expression for its width ($\hbar = c = 1$):

$$\Gamma_{rl} = \alpha_\mu^{(d)}(i \to f) \cdot \frac{8\pi(L+1)}{[(2L+1)!!]^2} \omega^{2L+1} b(EL; 0 \to \omega). \tag{1}$$

Here $L$ is the multipole order, $\omega$ is the transition energy, with $b(EL; 0 \to \omega)$ being the strength function for the nuclear excitation [8]. Then, $\alpha_\mu^{(d)}(i \to f)$ is the MCC for the muonic transition from level $i$ to $f$. Superscript $d$ underlines the fact that both muonic states belong to the discrete spectrum as distinct to the traditional internal conversion, when an electron (or a muon) transfers into the continuum. As a consequence, the discrete MCC become energy dimensional due to the other normalization of the wavefunction of the corresponding muonic state. Discrete MCC were earlier used by Zaretsky and Karpeshin. They predicted the previously mentioned effect of emission of muonic X-rays from the heavy fragment as a result of muonic promotion to the $2p$ state via internal conversion. To calculate $\alpha_\mu^{(d)}(i \to f)$ a number of programs was used from the RAINE set [9] intended for relativistic calculations of atomic structure, and modified for calculation of MCC. Finite nuclear extent, vacuum polarization and electronic screening are included along with relativistic effects. The nuclear electromagnetic strength functions have been calculated by means of the formula

$$b(EL; 0 \to \omega) = \sum_g B(EL; 0 \to \omega) \cdot \frac{\Delta/2\pi}{(\omega - \omega_g)^2 + (\Delta/2)^2}, \tag{2}$$

where $\omega_g$ the energy of a phonon state $g$, $\Delta$ being the averaging parameter. The procedure for calculating the strength functions is described in detail by Soloviev et al. [8]. We have used in the calculations the value of $\Delta = 300 - 500$ keV. At the excitation energy of $6 - 10$ MeV the density of the one-phonon states with given $L$ is about 100 per MeV. The values of $\Delta$ used are fairly large to smooth away fluctuations in $b(EL; 0 \to \omega)$ produced by particular highly excited states $\omega_g$. On the other hand, such values of $\Delta$ are fairly small so as not to distort the average value of $b(EL; 0 \to \omega)$ at the given excitation energy.

The radiationless transition probability per muonic atom is given by the branching ratio

$$W_{rl} = \Gamma_{rl}/(\Gamma_{rl} + \Gamma_\gamma^{(i)}), \tag{3}$$

where $\Gamma_\gamma^{(i)}$ is the radiative width of the muonic state $i$. These widths were calculated allowing for the relativistic effects, analogously to $\alpha_\mu^{(d)}$, by using the Dirac wave functions.

Table1. Calculated radiationless transition probabilities (percent) in comparison with the experimental data. $\langle 2p - 1s \rangle$ is averaging over the fine-structure components. Designations SC, VC show application of the surface or volume nuclear current models, respectively, for the sake of calculation of MCC

|  | $\langle 2p - 1s \rangle$ | $2p_{1/2} - 1s$ | $2p_{3/2} - 1s$ | $3p - 1s$ | $3d - 1s$ |
|---|---|---|---|---|---|
| Experiment [7] | 26.2±2.6 | 21.6±3.2(1.6) | 31.1±2.8(1.3) | 88.9±4.3 | 12.8±1.4 |
| Zaretsky & Novikov with σ from [10] | 22.4 | 21.1 | 24.2 | 64.7 |  |
| Zaretsky & Novikov with σ from [11] | 29.8 | 28.4 | 32.0 | 68.5 |  |
| Teller & Wess [3] | 20.7 | 20.0 | 21.7 | 59 | 9.6 |
| Karpeshin & Nesterenko [4] | $11 - 15^{SC}$ $19 - 26^{VC}$ |  |  | $55 - 65^{SC}$ $57 - 69^{VC}$ | $19 - 24^{SC}$ $25 - 32^{VC}$ |

## 2. Results and discussion

Note that a statement can be noted saying that the radiative widths of *np* muonic levels for *n* > 2 are mainly determined from the transition to the 1*s* state. However, the probabilities of the transitions to the 2*s* state turned out to be approximately equal. It is the finite nuclear extent which produces such an effect. For these transitions the strength function is mainly determined from the giant electric dipole resonance (GDR). However, their energies are smaller than the energy of the top of the GDR. This refers especially to the 2*p*→1*s* transitions but to a lesser degree also applies to the 3*p*–1*s* transitions in which the energy falls on a slope of the GDR. Under these circumstances, the collective mode of nuclear motion is not manifested so clearly. As a result, the strength function might not be calculated as reliable.

For this talk, I revised the calculations for the *E*1 nonradiative transitions in $^{238}$U by making use of experimental photoexcitation cross-sections from the paper by Caldwell et al. [11]. Experimental data are perfectly fitted by two-humpered GDR. The formula for $\Gamma_{rl}$ is factorized into MCC, the cross-section and geometric factor. The formula is very simple and universal for each kind and order of multipolarity. Indeed, cross-section can be expressed in terms of the strength function as follows:

$$\sigma_L = 8\alpha\pi^3 \cdot \frac{(L+1)\omega^{2L-1}}{L[(2L+1)!!]^2} b(EL; 0 \to \omega). \quad (4)$$

Making use of (4) in (1), one arrives at the following expression:

$$\Gamma_{rl} = \alpha_\mu^{(d)}(0 \to \omega) \cdot \frac{\omega^2}{\pi^2} \sigma_L. \quad (5)$$

The results of calculation are presented in Table 2. As above, there is a good agreement for the 2*p*–1*s* transition. However, there remains drastic contradiction between the theoretical and experimental probabilities in the case of the 3*p*–1*s* transition. A particular attention was paid to the ratio of the 3*p*–1*s* nonradiative-to-radiative probabilities. It can be clearly seen from Table 2 that in the case of the 3*p*–1*s* transitions, the experimental nonradiative probability is by an order of magnitude as high as the theoretical one. In the case of the $2p$–$1s$ transition, the agreement with experiment is good.

Table 2. Comparison of theory, based on using experimental cross-sections [11], with experiment [7] with respect to the radiationless transition probabilities $W_{rl}$ and the ratio of the nonradiative-to-radiative transition widths in comparison with the experimental data.

| Transition | Energy, MeV | $W_{rl}$, % | | $\Gamma_{rl}/\Gamma_{\gamma}^{(i)}$ | |
|---|---|---|---|---|---|
| | | theory | experiment | theory | experiment |
| $2p_{3/2}$–$1s$ | 6.5 | 31.6 | 31.1±2.8 | 0.46 | 0.45 |
| $3p_{3/2}$–$1s$ | 9.5 | 66 | 88.9±4.3 | 1.92 | 15 |

## 3. Conclusion

Prompt fission provides multilateral information about fission dynamics. We understand a lot of data concerning the $2p$–$1s$ nonradiative transitions. Argumentation of fission barrier and suppression of the fission mode is of great interest, as well as properties of the fragments.

At the same time, the most intriguing seems the contradiction between theory and experiment for the $3p$–$1s$ transition. The experimental non-radiative width is 7 times higher than the theoretical one. It seems highly heuristic to involve the processes of throwing the muon back. Otherwise, it can be the fine structure of the GDR. In this case, the muon turns out to be a unique tool for investigating GDR structure with high resolution by 100% definite multipolarity monochromatic photons. Note that fission channel is suppressed by about an order of magnitude in the case of the $2p$–$1s$ radiationless transition. But experimental cross-section which is used in the calculations involves the photofission cross-section fully. How does this affect probability? And in the case of the $3p$–$1s$ transition, the suppression is not confirmed experimentally.

The simplest conjecture concerning the $3p$–$1s$ line broadening is based on the experimental non-radiative transition probability. As that is 15 times greater than the radiative one (see Table 2), therefore, broadening achieves as much as a factor of 15. That means that it increases from about 0.2 to 3 keV. At the same time, however, many details and questions remain unclear. There is also level doubling due to the non-radiative interaction, with the related broadening of the second radiative component within MeV scale. Moreover, the nucleus gets excited, properly speaking, not in the $3p$–$1s$ transition, but rather in the preceding cascade transition to this state, like $4d$–$3p$, even $3d$–$3p$ (virtually) or similar. Correspondingly, some missing intensities should manifest themselves in these transitions. The basis for such consideration is laid in Ref. [12] using an example of electronic atoms of $^{229}$Th.

Regarding the fine structure of the giant resonances, all what is said above concerning their fine structure especially refers to the $3d$–$2p$ nonradiative transition accompanied with excitation of LEOR. It is very appropriate to perform independent measurements on the $^{235}$U isotope and other actinides. Undoubtedly, continuation of this research will yield in new unexpected discoveries.